\documentclass{aip-book}

\usepackage{graphicx,hologo,mwe,url}
\usepackage{array,booktabs,threeparttable} 
\usepackage{caption}

\usepackage[utf8]{inputenc}
\usepackage[american]{babel} 
\usepackage{csquotes} 

\usepackage[style=apa, backend=biber]{biblatex}
\usepackage{amsmath}
\usepackage{amsthm}
\usepackage{hyperref}

\theoremstyle{definition}

\begin{document}
\chapter{
Curiosity and Metacognition: Towards a Unified Framework for Learning and Education in the Age of AI
} 


\chapterauthor{Chloé Desvaux}{Inria Center of University of Bordeaux}{}{chloe.desvaux@inria.fr}
\chapterauthor{Rania Abdelghani}{Hector Institute, University of Tübingen}{}{rania.abdelghani@uni-tuebingen.de}
\chapterauthor{Pierre-Yves Oudeyer*}{Inria Center of University of Bordeaux}{}{pierre-yves.oudeyer@inria.fr}
\chapterauthor{Hélène Sauzéon}{Inria Center of University of Bordeaux}{}{helene.sauzeon@inria.fr}

\begin{abstract}
This chapter examines the relationship between curiosity and metacognition as critical drivers of autonomous and self-regulated learning. We synthesize recent research to propose a unified framework integrating behavioral, computational, and psychoeducational dimensions, arguing that curiosity—the intrinsic drive to acquire new knowledge—relies fundamentally on metacognitive monitoring and control. From an educational perspective, we evaluate interventions designed to enhance curiosity in classroom settings. While promising, our review indicates that these interventions yield mixed results, often proving differentially effective for struggling learners, thereby underscoring the necessity for approaches tailored to individual profiles. Finally, we address the paradigm shift introduced by Generative AI. While Large Language Models (LLMs) offer unprecedented scalability for personalized inquiry, we argue that their default interaction modes pose significant risks to the dynamics of curiosity-driven learning. To mitigate these challenges, we review strategies to transform AI from a potential cognitive shortcut into a powerful partner for sustained epistemic development.
\end{abstract}


\section{Introduction}
Curiosity—defined as the intrinsic drive to seek out new knowledge for the sheer joy of learning—has been increasingly recognized as a cornerstone of children’s autonomous learning, academic success \parencite{shahEarlyChildhoodCuriosity2018}, and well-being \parencite{oecdEarlyLearningChild2020}. But what makes curiosity so pivotal in cognitive development? The answer lies in its deep connection with metacognition, the ability to "think about thinking" \parencite{flavell1979metacognition}. Metacognition involves monitoring and regulating one’s own cognitive processes, and it is this interplay that amplifies curiosity’s impact on learning.
Together, curiosity and metacognition form a critical cognitive, emotional and social skill set that educators are encouraged to nurture from an early age \parencite{jiroutChildrensScientificCuriosity2012, petersonSupportingCuriositySchools2020}. Yet, despite its importance, curiosity remains surprisingly rare in classrooms \parencite{engelCuriosityVanishing2009, engelChildrensNeedKnow2011}. 
Consider this example: In a primary school classroom, a teacher introduces a simple experiment: placing an ice cube on a plate and asking students, ``What will happen if we leave it here all morning? Why?" Hands shoot up, hypotheses abound (``It will melt!", ``No, it will disappear!"), and unexpected questions emerge (``What if we put salt on it?", ``Does sugar melt too?"). The teacher seizes the moment, encouraging students to design their own mini-experiments to test their ideas. Here, curiosity is sparked by uncertainty and open-ended inquiry, while metacognition is activated as students realize their own knowledge-gap, explore, plan, observe, and reflect on their reasoning. This scenario, however, is all too rare in today’s classrooms, where standardized curricula, time constraints and rigid lesson plans often overshadow such exploratory moments, stifling curiosity's potential to thrive.

In response, researchers have turned to innovative solutions. Some studies have examined digital interventions designed to spark curiosity, for instance by introducing uncertainty \parencite{lamninaDevelopingThirstKnowledge2019} or by providing external cues such as pieces of novel or conflicting information to explore or starters for questions that could expand the understanding of material at hand \parencite{fitzgibbonCounterfactualCuriosityPreschool2019}—features that have been shown to induce curiosity states and exploratory behaviors in young children \parencite{liquinExplanationseekingCuriosityChildhood2020}. Others focus on strengthening the metacognitive strategies that underpin curiosity \parencite{jiroutCuriosityChildrenAges2024}. Now, with the rise of generative AI that pose new educational challenges, there are also new emerging opportunities to promote curiosity-driven education—particularly by leveraging the synergistic relationship between curiosity and metacognition ~\parencite{Abdelghani2023GPT-3-drivenSkills}.

This chapter offers a three-part overview of the strong ties between curiosity and metacognition, and their educational implications in the age of AI, drawing on: 
\begin{itemize}
\item  \textbf{Behavioral and computational approaches}—How do the links between metacognition and curiosity shape learning performance and experience?
\item  \textbf{Psychoeducational perspectives}—How can we harness the bidirectional relationship between curiosity and metacognition to create targeted interventions and enhance learning? 
\item  \textbf{The shift with generaive AI}—How can generative AI change curiosity–metacognition interactions in educational settings? And how can it be leveraged to strengthen educational interventions?

\end{itemize}
\section{Curiosity and Metacognition: A Dynamic Duo for Learning}
\label{sec:theory}
Modern research on curiosity, though still young, is evolving rapidly. Once viewed as a single, isolated trait, curiosity is now understood as a dynamic, multi-layered process—one that bridges basic learning mechanisms with higher-order cognitive functions, particularly the interplay between learning and metacognitive processing, at various timescales ranging from seconds to years.

\subsection{From single perspective to multi-perspective accounts of curiosity}
Among modern theories of curiosity, four single-component perspectives stand out, each offering a distinct yet complementary lens: 1) The \textit{learning} account – viewing curiosity as a driver of knowledge acquisition ; 2) The \textit{metacognition} account – framing curiosity as a process intertwined with self-regulated thinking; 3) The \textit{uncertainty regulation} account – understanding curiosity as a mechanism for managing or reducing uncertainty. and ; 4) The \textit{emotion} account – interpreting curiosity as an affective state that motivates exploration. These accounts coexist in the literature, each enriching our understanding of curiosity without excluding the others.

\begin{figure}
\centering
\includegraphics[width=1\linewidth]{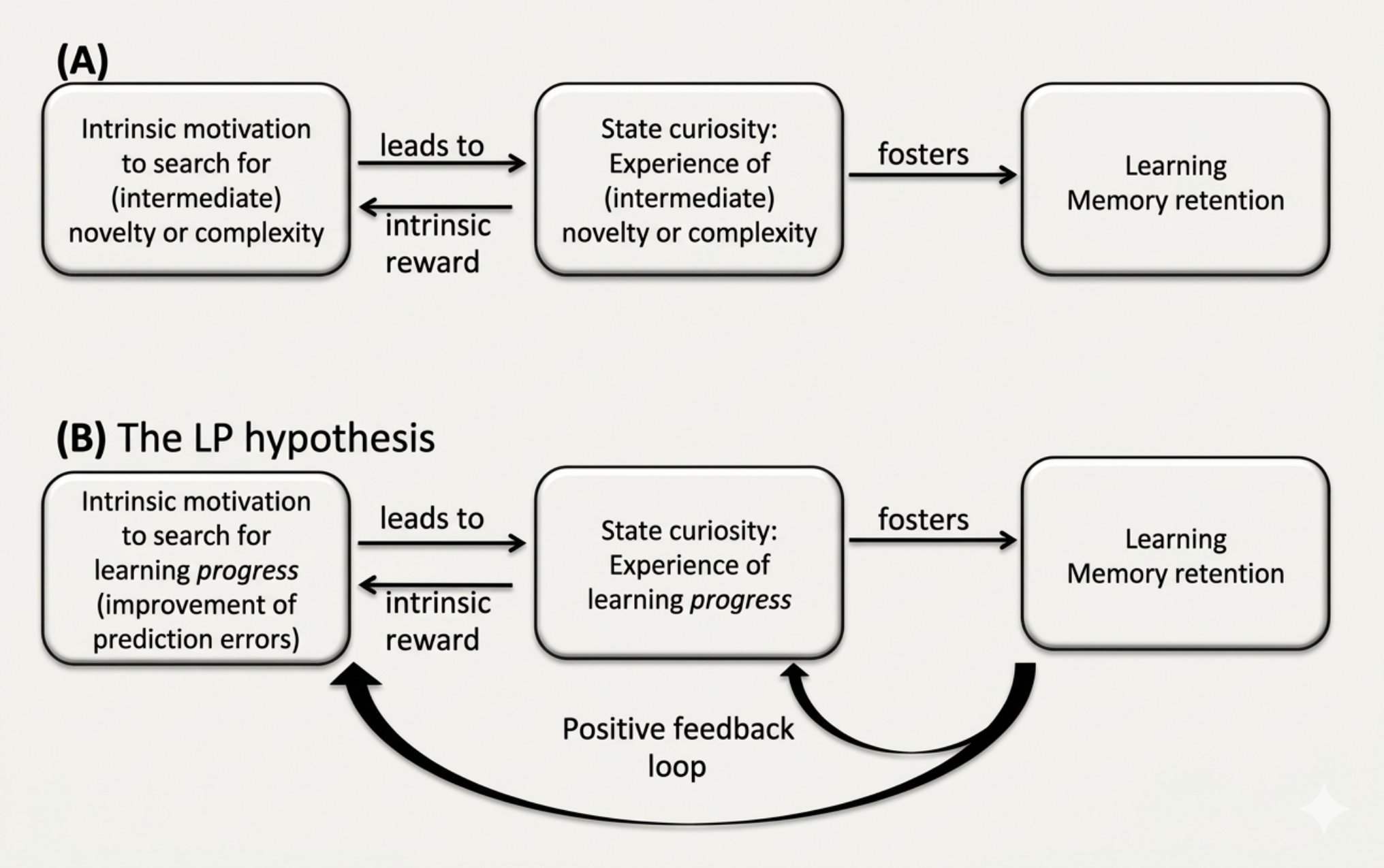}
\caption{\footnotesize\label{fig:lp}  \centering \textbf{(A) }A one directional causal relationship between state curiosity and learning, as conceptualize in some studies. \textbf{(B)} The LP hypothesis where learning progress itself, measured as the change in prediction or control errors, can be intrinsically rewarding such as a positive feedback loop between state curiosity and learning (B). Put simply, self-detection of a new learning brings positive feedback associated with curiosity experience, leading in turn to continue exploring and learning further: that is a core engine of open-ended learning. (From~\parencite{oudeyer2016intrinsic})}
\end{figure}

The curiosity\textbf-as-{learning perspective} frames curiosity as a form of intrinsic motivation—engagement in activities driven by inherent satisfaction, such as enjoyment, interest, or a sense of competence and autonomy, rather than external rewards ~\parencite{ryanIntrinsicExtrinsicMotivations2000}. Unlike extrinsic motivation, which relies on external outcomes (e.g., praise or grades), intrinsic motivation includes phenomena like flow states during skilled performance ~\parencite{wong1991motivation}. As a subset of intrinsic motivation, curiosity processes are specifically oriented as drivers or spontaneous exploration, novelty-seeking, and knowledge acquisition ~\parencite{oudeyer2007intrinsic}. It has been described as "the impulse toward better cognition" ~\parencite{kashdan2012whether}  or the pursuit of novel sensations and experiences for learning or non-instrumental information-seeking ~\parencite{james1890principles, kiddPsychologyNeuroscienceCuriosity2015, loewensteinPsychologyCuriosityReview1994}. In real-world contexts, exploratory behaviors often blend intrinsic and extrinsic motivations. For instance, in education, activities may engage students' natural curiosity while also serving instrumental goals, such as exam preparation ~\parencite{deci2001extrinsic, murayamaProcessAccountCuriosity2019}. The learning account employs computational and empirical approaches, including behavioral indicators (e.g., information-seeking actions, exploration time ~\parencite{Abdelghani2022ConversationalChildren, evans2023investigating}), physiological measures (e.g., pupil dilation, gaze patterns ~\parencite{baranes2015eye}), and neuroimaging \parencite{gruber2014states, kangWickCandleLearning2009}. These multi-method approaches reveal how curiosity shapes attention, memory, and learning outcomes, offering opportunities to design educational technologies with genuine added value \parencite{gottlieb2018towards, oudeyer2016intrinsic}.
A key example is the\textit{ Learning Progress (LP)} hypothesis, which highlights the positive feedback loop between intrinsic motivation, curiosity states, and self-evaluation of learning gains (see also the ``information-as-reward hypothesis" ~\parencite{marvin2016curiosity}), as illustrated in \autoref{fig:lp}.

A closely related perspective is curiosity as \textbf{uncertainty regulation}, rooted in Berlyne’s conflict theory, which defines curiosity as an aversive motivation driving individuals to seek knowledge to resolve the experienced uncertainty \parencite{berlyne1966curiosity}. Accordingly, curiosity is satisfied when the pursued knowledge is obtained, and its intensity depends on the initial level of uncertainty: the greater the conflict or ambiguity in a situation, the stronger the curiosity it evokes.
Loewenstein’s ``knowledge gap" theory further refines this idea, framing curiosity as a cognitive deprivation arising when individuals perceive a discrepancy between their current and desired knowledge states ~\parencite{loewensteinPsychologyCuriosityReview1994}. However, curiosity is not linear: when the knowledge gap is too large (e.g., the information is entirely unknown) or too small (e.g., the answer is already certain), motivation to seek information declines. For example, ~\cite{kangWickCandleLearning2009} demonstrated that individuals are least curious when they lack any clues about an answer or are overly confident in their response. Similar patterns have been observed in children, particularly in tasks involving spatial orientation and memory \parencite{sivashankar2024beneficial}.
An optimal level of uncertainty is therefore necessary to trigger information-seeking behavior. These mechanisms are believed to underpin children’s curiosity-driven exploration from as early as age 4 ~\parencite{molinaro2023multifaceted}. Loewenstein’s theory also explains why curiosity is often sparked by stimuli perceived as surprising, novel, or inconsistent with prior beliefs ~\parencite{barto2013novelty, oudeyer2007intrinsic, poli2020infants}. It should be noted that Loewenstein's theory is in accordance with the first two accounts.

The \textbf{metacognitive account} of curiosity is grounded in research distinguishing between metacognitive knowledge (declarative metacognition) and metacognitive experiences (procedural metacognition) \parencite{flavell1979metacognition}.

\begin{figure}[H]
\centering
\includegraphics[width=1\linewidth]{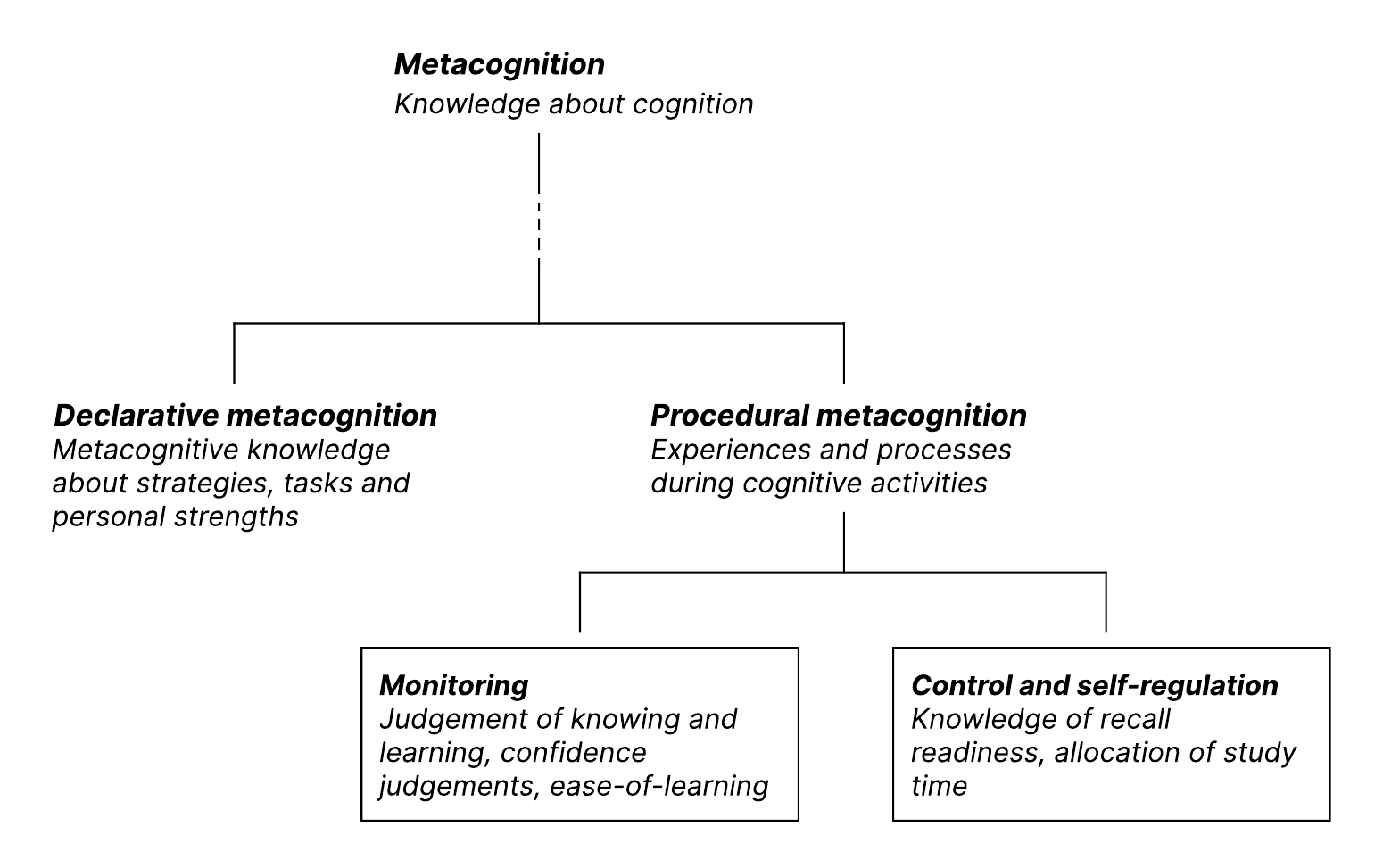}
\caption{\footnotesize\label{fig:mc}  \centering Overview of metacognition in terms of knowledge and process (adapted from \textcite{schneider2013procedural})}
\end{figure} 
Metacognitive knowledge refers to an individual’s awareness of their cognitive strengths, tasks, and strategies (e.g., recognizing a stronger aptitude for mathematics than for arts). In contrast, procedural metacognition involves the conscious experiences and processes that occur during cognitive activities, such as the feeling of not understanding something or the sense of knowing ~\parencite{schneider2013procedural} \autoref{fig:mc}.
Within this account, curiosity is understood as both a state and a metacognitive feeling—a subjective experience that signals cognitive states and learning opportunities ~\parencite{goupilCuriosityMetacognitiveFeeling2023, metcalfeEpistemicCuriosityRegion2020, wade2019role}. Unlike stable personality traits (e.g., those measured by the Five-Dimensional Curiosity Scale ~\parencite{kashdan2020five}), this account emphasizes transient experiences of curiosity, such as detecting knowledge gaps or prediction errors. It also highlights the role of declarative knowledge about curiosity in driving curious behaviors. For example, Post and Walma van der Molen demonstrated how children’s perceptions and beliefs about curiosity (including their attitudes and images of curiosity) influence the actual expression of curious behaviors~\parencite{Post2019DevelopmentQuestionnaire, post2020effects}.

Finally, curiosity as \textbf{an emotion or affective state}—the desire to learn—is a fast, multicomponent phenomenon involving elicitation (e.g. arousal triggered by the detection of information gaps) and response (characterized by active, goal-directed exploration) phases ~\parencite{connelly2011applying, nerantzaki2021epistemic, silvia2005interesting, silvia2008interest, silvia2010confusion}. Appraisal theories suggest curiosity arises from automatic environmental evaluations, triggering physiological, motor, and motivational responses~\parencite{scherer2019emotion}. While its classification as an emotion is debated, studies confirm it involves subjective feelings ~\parencite{tang2022differences} , bodily expressions ~\parencite{lyu2022spontaneous}  and exploratory tendencies  ~\parencite{poli2024curiosity}. Research on interest and epistemic emotions further supports this view, emphasizing affective dynamics like surprise, confusion, enjoyment, frustration, and boredom ~\parencite{muis2015role}.

As a result of these multiple accounts, curiosity is increasingly understood as a dynamic, multicomponent process—integrating attentional, motivational, emotional, cognitive, and metacognitive mechanisms to guide exploration and learning. For example, ~\textcite{erdemli2025integrative} propose an integrative framework of epistemic curiosity, combining information-gap, emotion, and reward-anticipation accounts. Here, curiosity is framed as an emotion triggered by appraising a knowledge gap as relevant, valenced, novel, and manageable. This appraisal assigns reward value to the object of curiosity, motivating knowledge-seeking (``wanting knowledge"). Once satisfied, future expectations are updated via affective prediction errors (``liking and learning"). However, if relevance is high but coping potential low, the gap generates anxiety and leads to information avoidance.
While this model is supported by behavioral studies and emerging neuroscience, it requires further validation. ~\textcite{erdemli2025integrative} recommend testing the appraisal of informational gaps and formal validation through computational approaches. Future research should elucidate the interactions among curiosity’s components. Below, we explore integrative advances linking intrinsically motivated learning, uncertainty, and metacognitive accounts.

\subsection{A unifying view of curiosity including intrinsic rewards and metacognition}
Both empirical and computational approaches ground the modern view of curiosity where metacognition aspects are integrated.

\begin{figure}
\centering
\includegraphics[width=1\linewidth]{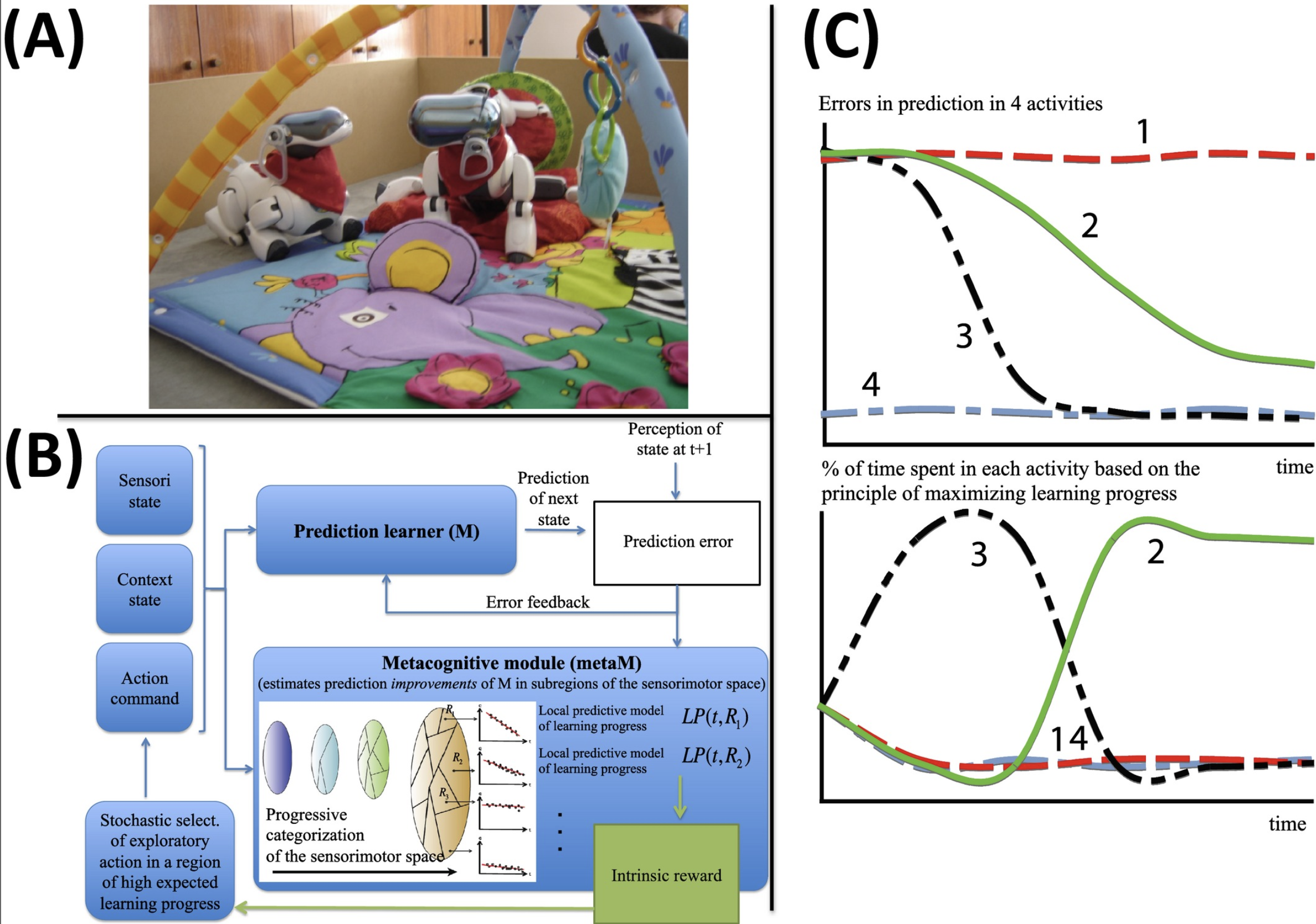}
\caption{\footnotesize \label{fig:comp} \centering  
The playground experiment ~\parencite{oudeyer2016evolution}. \textbf{(A)} The learning context. \textbf{(B)} Computational architecture for curiosity-driven exploration used in robotic modeling experiments (here architecture for modeling knowledge-based intrinsically motivated exploration mechanisms). A prediction module (M) learns to anticipate the sensory consequences of a robot’s actions (e.g., predicting visual or proprioceptive outcomes of a leg movement via a neural network). A metacognitive module tracks how prediction errors evolve across the sensorimotor space, estimating error reduction (e.g., for leg movements in specific environmental regions). These estimates generate intrinsic rewards, proportional to the decrease in prediction errors, which guide action selection in a reinforcement-learning framework ~\parencite{oudeyer2007intrinsic}. The system preferentially explores actions with high estimated learning progress, but does so probabilistically to allow discovery of new rewarding activities. Since sensorimotor flow is not pre-segmented, the system progressively categorizes it into distinct regions, modeling the incremental formation of cognitive activity/task categories. \textbf{(C)} Illustration of a self-organized developmental sequence where the robot automatically identifies, categorizes, and shifts from simple to more complex learning experiences. Figure adapted with permission  ~\parencite{gottlieb2013information}
}
\end{figure}

\textbf{Computational models} provide powerful frameworks for testing curiosity theories, simulating mechanisms, and generating testable predictions. For instance, they have replicated developmental trajectories (\textit{e.g.}, vocal development ~\parencite{moulin2014self}) and predicted exploratory behaviors in humans and animals ~\parencite{gordon2014emergent}.
A notable example is multi-component computational models of curiosity, which distinguish from a primary learning system (knowledge/skill acquisition) to a metacognitive system (monitoring learning progress and generating intrinsic rewards) ~\parencite{baldassarre2012intrinsically, oudeyer2018computational}. This model explains how curiosity adaptively focuses on learnable challenges while avoiding trivial or overly complex tasks ~\parencite{oudeyer2018computational, tenHumansMonitorLearning2021} thanks to a dual-loop architecture : a Low-level one for interaction with the environment and model updating and a High-level (metacognitive) one for tracking learning progress, computing intrinsic rewards, and selecting experiments.
~\textcite{oudeyer2016evolution} illustrated this dual system in robot sensorimotor development (\autoref{fig:comp}) where a robot learner probabilistically selects experiences according to their potential for reducing uncertainty, that is, for learning progress.

\begin{itemize}
    \item \textit{Undirected exploration} (sometimes random or uniform) in which interest is uniformly distributed across the choice space (all learning situations are equally interesting).
    \item \textit{Directed exploration} (interestingness scaled with the agent’s abilities) including both \textbf{Knowledge-based intrinsically motivated exploration} where exploration is guided to improve a world model) and \textbf{Competence-Based Intrinsically Motivated exploration} where exploration happens by iterative self-generation and self-pursuit of new yet learnable goals.
\end{itemize}

Competence-based exploration, also referred to as 'autotelic' exploration or autotelic curiosity  \parencite{colas2022autotelic}, involves individuals generating and pursuing self-selected goals. This form of curiosity relies on higher-level cognitive functions, such as the ability to imagine abstract and novel goals—a capacity critical for creative discoveries and open-ended lifelong learning. Children often achieve this through language compositionality, using it as a tool to envision new situations during play \parencite{piaget2005language, vygotsky2012thought}. They also formulate divergent, curiosity-driven questions that yield informational gains through social interactions \parencite{Abdelghani2022ConversationalChildren, Abdelghani2023GPT-3-drivenSkills, Alaimi2020PedagogicalChildren}.
The IMAGINE architecture \parencite{colas2020language}, a deep reinforcement learning framework driven by intrinsic motivation, models this capacity within the dual-system model of curiosity. Like children, imaginative agents benefit from interactions with a social or metacognitive partner who provides linguistic descriptions of environmental interactions, such as navigational actions to pursue (e.g. "go bottom left"), or object manipulations (e.g. "grasp any blue thing") . These descriptions are interpreted as goals, enabling agents to learn to achieve them. Moreover, these descriptions can be compositionally recombined to enable the curious agents to self-generate entirely new goals beyond previously attained ones—a process termed compositional imagination (e.g. "grasp red cat"). This architecture demonstrates how generalization and exploration improve in agents equipped with these features, compared to those without. The linguistic properties of goal imagination (akin to inner language) and social/linguistic interactions are key enablers of these outcomes, fostering metacognitive development through self-reflective qualities \parencite{flavell1979metacognition}.
Complementing this, the MAGELLAN architecture \parencite{gaven2025magellan} goes beyond imagining new goals by adding a metacognitive layer to strategically select goals according to their predicted learning progress. MAGELLAN is thus derived from the dual-system model and the LP hypothesis, and computationnally illustrates the role language can play in curiosity-driven autotelic exploration. By leveraging large language models (LLMs) as providers of declarative knowledge, MAGELLAN enables agents to manage their own learning by tracking their skills and growth potential (monitoring of learning) while using this data to select useful and achievable goals (control). This confers the two core properties of metacognitive processing—monitoring and control—to the agent, while also equipping it with evolving metacognitive knowledge to shape its skill trajectory. 

From \textbf{an empirical perspective}, recent neuroscience studies highlight distinct neural networks activated during curiosity-driven tasks. The PACE model for memory performance~\parencite{gruberHowCuriosityEnhances2019} identifies four cyclically interacting components: \textbf{P}rediction (Generating expectations), \textbf{A}ppraisal (Evaluating information value), \textbf{C}uriosity state (Subjective experience), and \textbf{E}xploration (Information-seeking to resolve prediction errors).
~\textcite{gruber2014states} found that high curiosity during anticipation engages the reward system more strongly, with increased functional connectivity between the reward system and hippocampus. This suggests dopamine release from the midbrain enhances hippocampal memory encoding. \textcite{duan2020effect} contrasted intrinsic and extrinsic motivation, showing that curiosity-related anticipation activates the reward system, middle temporal gyrus, and inferior parietal lobule—regions linked to semantic access and uncertainty resolution. Researchers interpreted these additional activations as evidence of metacognitive monitoring, wherein the brain assesses existing knowledge and identifies information gaps to regulate later learning. During the answer phase, curiosity increased activity in the reward system and fronto-parietal attention networks, while extrinsic rewards led to deactivation in parietal regions. These findings indicate that curiosity’s memory benefits stem from enhanced attentional resources for cognitive monitoring and control.

In a related way, Murayama’s reward-learning framework~\parencite{murayamaProcessAccountCuriosity2019} describes curiosity as a cycle where knowledge gaps trigger exploration, leading to intrinsically rewarding knowledge acquisition. This aligns with the LP hypothesis ~\parencite{oudeyer2016intrinsic}, emphasizing curiosity’s interplay with metacognition: curiosity states shape metacognitive processing, and vice versa. Curiosity acts as a trigger for metacognitive experiences of knowledge gaps, while metacognitive processing guides curiosity-driven learning by predicting learning gains and assessing information value.

Empirical evidence shows that metacognitive monitoring of knowledge gaps is central to curiosity-driven learning. For example, in trivia tasks, curiosity levels are higher before knowing an answer than after~\parencite{fastrich2018role}. Similarly, curiosity peaks when metacognitive judgments of learning are high (\textit{e.g.}, on the verge of progress) ~\parencite{metcalfe2024curiosity}. Free-choice paradigms also confirm that humans use metacognitive monitoring of learning progress to prioritize goals ~\parencite{leonard2023young,poli2024curiosity,sayali2023learning, tenHumansMonitorLearning2021}. For instance,~\textcite{tenHumansMonitorLearning2021} showed that individuals rely on competence information to avoid easy tasks, with models including a learning-progress component best fitting task selection data.
 ~\textcite{kimCuriosityCanInfluence2025} found that curiosity states were higher for incorrect answers than correct ones in trivia tasks, linking curiosity to metacognitive monitoring of the ``known-unknown" distinction. In this study, authors further show that these high curiosity states are further associated with meacognitive control namely by influencing decisions to skip questions rather than risk incorrect answers.
Beyond monitoring and control, metacognitive knowledge about curiosity shapes curiosity-driven learning. The foundations of such metacognitive knowledge may emerge as early as age 3 ~\parencite{broesch2017learning}.~\textcite{jeong2025young} found that, compared to 4- to 5-year-olds, 6-year-olds’ improved ability to formulate curious questions is associated with enhanced metacognitive knowledge. Among which knowledge about malleability of personal skills in asking questions (\textit{e.g.} ``Do you think you are better at the question game now? Or do you think you would be better at it when you are 10 years old?'' or ``There are two kids playing. One is trying the question game for the first time, and the other kid has played it 10 times before. Who do you think will play the game better?''). This goes for curiosity-related processes more generally: understanding curiosity as a malleable skill that enhances lifelong learning amplifies the expression of curiosity behaviors.
Additionally, metacognitive procedures—such as strategic planning, uncertainty management, and the leveraging of social resources—are essential for driving curious behavior. This involves several critical facets of metacognitive knowledge such as fostering personal interest ~\parencite{hidi2019interest,keller1987strategies, renninger2015power, sinha2017new}, mastering active investigation and open exploration strategies like hypothesis testing, and embracing challenges beyond one’s current abilities ~\parencite{kapur2015learning,kiddPsychologyNeuroscienceCuriosity2015}. Furthermore, recognizing that uncertainty is socially legitimate ~\parencite{jordan2014managing} and knowing that curiosity evokes a state of pleasurable excitement ~\parencite{berlyne1966curiosity} constitute vital metacognitive knowledge for sustaining these behaviors in daily life. For these reasons, educational environments—whether at home or in school—that frame curiosity as a socially desirable process, emphasizing its lifelong benefits for learning and skill development, encourage the expression of curious behaviors~\parencite{goudeau2023unequal,goudeau2024causes, petersonSupportingCuriositySchools2020}.
Lastly, because learning does not happen only at the individual level, social metacognitive functions also play a significant role in learning dynamics: awareness of others' knowledge, conflicting beliefs or shared conceptions of a problem are essential abilities to develop ~\parencite{sinha2017new}.

Finally, the interplay between curiosity and metacognition unfolds across two broad dimensions. First, metacognitive processing involves actively assessing one’s knowledge gaps, evaluating the ability to meet challenges, and gauging the value of newly acquired information—while simultaneously tracking learning progress. This monitoring function works hand-in-hand with control processes, where learners formulate goals and deploy exploratory or inquiry-based strategies to steer their self-regulated learning.
Second, metacognitive knowledge about curiosity encompasses both an understanding of what curiosity entails—such as recognizing it as a malleable skill that fuels lifelong learning and holds social value—and a practical grasp of how curiosity operates. This includes mastering effective exploration strategies, navigating uncertainty, and leveraging social or environmental resources to deepen learning.

Together, these intertwined dimensions enable self-regulated and open-ended lifelong learning. The dual-system model of curiosity bridges seemingly disparate theories—such as learning progress as intrinsic reward, information-gap resolution, and metacognitive feelings—by demonstrating how they represent different operational modes of a single, underlying system. Looking ahead, future research could examines deeper into how these metacognitive dimensions evolve over time, shaping curiosity-driven learning behaviors across diverse contexts.


\section{School-level Interventions for Fostering Curiosity and their Limitations} 
\label{sec:interventions}
As discussed in Section \ref{sec:theory}, curiosity-driven learning is associated with more efficient and lasting learning outcomes, making it a valuable ability to foster in educational settings. However, despite its benefits, curiosity-driven learning remains underrepresented in many classrooms \parencite{engelChildrensNeedKnow2011}. In response, a growing body of research has examined school-based interventions aimed at promoting curiosity and related learning behaviors, sometimes under different labels such as inquiry-based learning or, in a more general sense, self-regulated learning.

Most of these interventions focus on metacognition as a primary medium for improvement. By developing students' metacognitive abilities—particularly monitoring their knowledge and progress, and controlling their learning processes—interventions aim to equip learners with lasting skills that enable them to engage in curiosity-driven learning autonomously. However, hands-on metacognitive training alone may not be sufficient to encourage curiosity behaviors in the classroom. In parallel, some researchers argue that the classroom climate must also be conducive to curiosity expression \parencite{jiroutChildrensScientificCuriosity2012, Post2018DoContext}. In this view, students' immediate school environment (sometimes defined as the microsystem \parencite{kimInterdisciplinaryReviewSelfRegulation2023, petersonSupportingCuriositySchools2020}), namely peers and teachers, must develop an understanding and acceptance of curiosity and its value in order to create an environment that encourages curious behaviors. These various levers for school-based interventions are depicted in figure~\ref{fig:school} below. 

\begin{figure}
\includegraphics[width=0.9\linewidth]{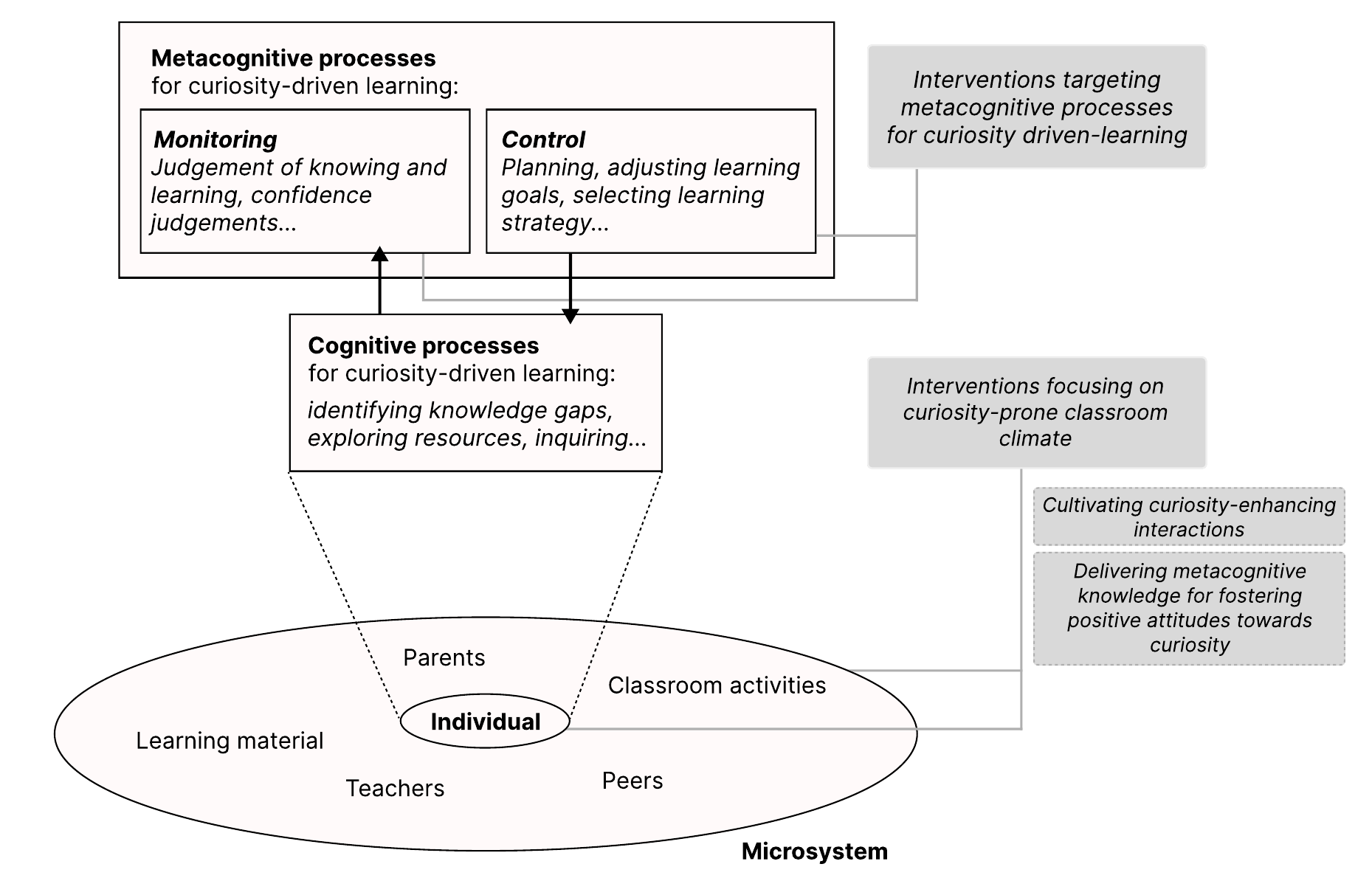}
\caption{\footnotesize\label{fig:school} Schematic representations of the domains targeted by school interventions for fostering curiosity in the classroom. Two main domains can be distinguished: (1) metacognitive processes relevant in curiosity-driven learning, including both monitoring and control and (2) promoting a favorable classroom climate for the expression of curiosity. School-based interventions with this objective can be categorized along complementary (sometimes overlapping) axes: delivering metacognitive knowledge about curiosity and its importance in learning and cultivating expression of curiosity within the microsystem (\textit{e.g.}interactions with peers and teachers) that relay positive attitudes towards curiosity.}
\end{figure} 

In this section, we review school-based interventions designed to promote curiosity-driven learning through two complementary approaches: (1) supporting students' metacognitive processing abilities (monitoring and control) for autonomous, curiosity-driven learning, and (2) enhancing classroom climate for fostering positive perceptions of curiosity among students, teachers, and peers, by delivering individual and social metacognitive knowledge on curiosity-driven learning.

\subsection{Supporting metacognitive abilities for curiosity-driven learning}
The main approaches to training curiosity-driven learning—or more broadly, self-regulated learning—involve scaffolding, supporting, or prompting students' engagement with metacognitive processes \parencite{dignathHowCanPrimary2008, dignathComponentsFosteringSelfregulated2008}. These processes can be broadly categorized into two key components: monitoring one's knowledge and learning progress, and exerting control over the learning process, including planning strategies, setting goals, and evaluating learning outcomes. In this subsection, we discuss school-based interventions that target these two fundamental metacognitive processing-related abilities to foster curiosity-driven learning behaviors in the classroom.

\subsubsection{Training monitoring skills for curiosity-driven learning processes}
\label{subsec:monitoring}
As mentioned in Section~\ref{sec:theory}, metacognitive monitoring (\textit{e.g.} feelings of knowing) can initiate curiosity states and exploratory behaviors as a way to fill in knowledge gaps \parencite{Muijs2020MetacognitionReview, ronfardQuestionaskingChildhoodReview2018}. Therefore, accurate metacognitive judgments are essential for acknowledging and pursuing novel learning goals, making them the starting point of any curiosity-driven learning process \parencite{murayamaProcessAccountCuriosity2019}.

Many school-based interventions have aimed to improve students' metacognitive monitoring accuracy by employing different methods, such as providing cues to support knowledge monitoring or implementing confidence ratings coupled with external feedback for improvement.
The first approach—the use of explicit cues to foster metacognitive monitoring—constitutes an effective method for making students aware of their missing knowledge, which is a crucial factor that motivates individuals to engage in information-seeking behaviors \parencite{metcalfeEpistemicCuriosityRegion2020}.
For instance, \textcite{Alaimi2020PedagogicalChildren} used a digital pedagogical agent to prompt primary school students to ask curiosity-driven questions about short educational texts. To make knowledge gaps salient, the agent provided specific semantic cues—pieces of information related to the text but likely unknown to students—and then prompted children to formulate a question about this new information. For example, after reading a text about living conditions on Mars, students might be provided with the cue ``\textit{using powerful spaceships}'' and asked to generate a question on that topic. Results showed that after this intervention, students were better at asking curiosity-driven questions autonomously. Using a similar task and setting, \textcite{Abdelghani2022ConversationalChildren} demonstrated the same pattern of results. Additionally, their intervention improved children's learning during an autonomous exploration task. In a later study, the authors \parencite{abdelghani:tel-04697786} used a digital metacognitive training intervention to support primary school children in identifying knowledge gaps and engaging in a curiosity-driven inquiry cycle. The training improved children's metacognitive sensitivity (their ability to accurately distinguish when they know or do not know an information) immediately after the intervention, though this effect did not persist at delayed post-testing. Importantly, the training intervention also enhanced children's ability to initiate curiosity-driven inquiries and generate questions.

Other interventions have aimed more specifically at improving metacognitive monitoring accuracy, particularly by training students to make confidence judgments and providing feedback on their accuracy. For instance, \textcite{proustCanMetacognitiveMonitoring2025} evaluated a metacognitive monitoring accuracy training intervention for elementary through high school children. In this school-based training, students made prospective and retrospective confidence judgments about their performance and received feedback comparing their predictions to their actual results. Results showed that regular use of this confidence rating tool, integrated into classroom activities, improved children's predictive monitoring accuracy.
However, no direct improvement was found on overall academic outcomes, suggesting that while the intervention successfully enhanced monitoring abilities, the translation to academic performance may require additional support or longer implementation periods.

\subsubsection{Supporting metacognitive control for self-regulated curiosity-driven learning}
The other aspect of metacognition that can be useful for autonomous and curiosity-driven learning processes is metacognitive control. As detailed in Section~\ref{sec:theory}, metacognitive control and regulation consists in using information from metacognitive monitoring in order to adapt learning behaviors accordingly \parencite{Griffin2013SupportingMonitoring}. While many school-based interventions focused on training or scaffolding monitoring abilities for learning performance, fewer have focused on explicitly supporting metacognitive control and regulation of learning—those are described below.

In a series of work on college and high school students, \textcite{Azevedo2008WhyHypermedia, Azevedo2009MetaTutorAM} used a digital learning platform in order to support students metacognitive regulatory processes in biology learning sessions. This tool displayed the content of the lesson accompanied by metacognitive prompts as a way to support self-regulated learning. For example, student were able to set their own learning subgoals, from a main learning goals set by the teacher. Other metacognitive scaffolding comprised explicitly selecting metacognitive strategies used by the students, for example, monitoring learning or planning the next learning steps. In these studies, authors showed that the use of the tool improved learning outcomes, and subjective reports of students suggest that they used more metacognitive strategies \parencite{Azevedo2008WhyHypermedia}, especially planning \parencite{Azevedo2009MetaTutorAM}.
In a mixed-methods study on elementary school children, \textcite{Hastuti2020TheStudents} implemented a guided inquiry-based learning intervention applied to mathematics lessons. Students were guided through goal setting, strategy planning and evaluation, as well as monitoring their progress throughout the learning process. Results of this study suggest that compared to a control group, students in the inquiry group made significantly more use of metacognitive processes during a mathematics post-test activity, as confirmed by interviews in which students were asked to elaborate on their problem-solving processes. Similarly, a case study by \textcite{Yildiz-Feyzioglu2013MonitoringPrompts} evaluated the effectiveness of a digital tool supporting self-regulated learning on middle school children. This tool consisted of lesson content on electricity, accompanied by metacognitive prompts and activities for self-regulated learning such as goal setting, monitoring, planning, and evaluation of learning. Qualitative results from this study suggest that the tool enabled students to regulate their goals, shifting from generic to more realistic and intrinsic goals after interacting with the tool. This shift in goal-setting quality suggests deeper engagement with the learning task and greater ownership of the learning process.
However, providing support, does not always guarantee improvement of performance. In a school-based intervention study on math lessons \parencite{fyfe2012effects}, metacognitive support was provided to elementary school children to enhance their metacognitive monitoring and control of strategies they used to solve the math problem. This was achieved by prompting students to explain their strategy choice and providing feedback on their performance and the appropriateness of their strategy for solving the exercise. Results from this study suggest that providing feedback on strategy choice and performance induced better procedural knowledge scores, but only for students with lower prior knowledge. Consistent with interventions presented in Subsection~\ref{subsec:monitoring}, this finding suggests that metacognitive scaffolding is most effective for lower-achievers, and does not always translate for all learner profiles.

\subsection{Promoting expressions of curiosity through enhancement of classroom climate and perceptions}
Complementary to the aforementioned approaches aimed at training specific metacognitive skills and strategies to promote curiosity-driven learning, some studies have focused on establishing supportive classroom environment and transmitting positive beliefs of curiosity in order to promote its expression. This approach aims at bridging the gap between hands-on training interventions and supporting students in developing a true understanding of the importance of curiosity in learning \parencite{Post2018DoContext}. This can be achieved through different methods, for example, by creating social interactions that promote expressions of curiosity with both peers and teachers, and by delivering specific metacognitive knowledge about curiosity and associated constructs to emphasize the importance and positive value of expressing curiosity in children's eyes.

\subsubsection{Peers and classroom climate fostering positive attitudes towards curiosity}
The promotion of a curiosity-prone classroom climate relies in large part on the social environment and sustained interaction with others \parencite{Post2018DoContext}, which allows for shaping individual behaviors \parencite{Bandura1986SocialTheory}, especially at the ages of formal schooling \parencite{petersonSupportingCuriositySchools2020}. In this light, some studies have attempted to promote curiosity by favoring positive interactions with a learning peer who expresses curiosity \parencite{Ceha2019ExpressionBehaviour, Gordon2015CanRobot}. 

\textcite{Gordon2015CanRobot} investigated whether children engaged in a reading task would be more curious when learning with a peer robot that expressed curiosity. They found that when the curious robot exhibited curiosity-related behaviors, such as enthusiasm for learning, suggesting novel exploration, and generating guesses, children showed significantly higher levels of free exploration in the task and uncertainty-seeking compared to those interacting with a non-curious robot. Consistent with this finding, \textcite{Ceha2019ExpressionBehaviour} used a robot expressing curiosity-driven behaviors such as question-asking as peer learner for a geology learning task. This study showed that participants could reliably recognize the robot's expression of curiosity, and that this recognition led to higher reported curiosity states as well as more curiosity behaviors. 

Another series of studies \parencite{Law2020CuriosityAgents, Lee2021CuriosityTeaching} focusing on learning-by-teaching in a digital environment positioned the student as the tutor to a less-knowledgeable virtual agent expressing curiosity. In these studies, elementary school students reported they viewed the agent as a good learner based on observable characteristics they perceived as positive, such as the agent's curiosity or its eagerness to learn. These positive perceptions were associated with students' feeling of competence as teachers, suggesting learning gains.

\subsubsection{Metacognitive knowledge and cognitive beliefs fostering positive attitudes towards curiosity}
Development of a classroom climate that encourages curiosity can also be achieved by delivering metacognitive knowledge and promoting positive cognitive beliefs about curiosity. For instance, in a recent study \parencite{abdelghani:tel-04697786}, authors evaluated a metacognitive intervention aimed at delivering metacognitive knowledge about curiosity and learning to students. Results from this study showed that the intervention significantly improved students' perceptions of curiosity.

Teachers also have a central role in shaping classroom climate through their interactions with students \parencite{petersonSupportingCuriositySchools2020}. For this reason, they tend to be the focus of numerous school interventions aimed at improving attitudes towards curiosity. Indeed, teachers' practices and pedagogical styles heavily influence students' expression of curiosity \parencite{Engel2009HowInquiry, Evans2023CuriosityClassrooms, Jirout2022DevelopmentProtocol}. For instance, a study by \textcite{Engel2009HowInquiry} showed that instructional goals set for teachers influenced the way they responded to their students' inquiry during a science lesson. Teachers instructed to complete the task were significantly more restrictive in their responses to students' inquiry and exploration behaviors compared to those instructed to learn more about the domain. 

With these considerations in mind, a series of studies \parencite{Post2021EffectsOrientations, VanAalderen-Smeets2015ImprovingDevelopment} evaluated the effectiveness of a school-based intervention specifically targeting teachers' attitudes towards inquiry-based learning. This intervention was successful in improving teachers' self-efficacy in their teaching activity as well as their enjoyment \parencite{VanAalderen-Smeets2015ImprovingDevelopment}. Consequently, when evaluating the effect of this intervention on promoting inquiry-based learning in the classroom, the results of their longitudinal study suggest successful improvement of students' attitudes and beliefs about curiosity \parencite{Post2021EffectsOrientations}. 
Similarly, another longitudinal study by \textcite{Alan2023ORE} evaluated a teacher-focused school intervention aimed at delivering knowledge about the importance of curiosity in learning and practices to foster it in students. Results of this study showed that the intervention significantly improved teachers' beliefs on curiosity. Quite interestingly, students in the intervention group also appeared to demonstrate more learning, as they performed better on a science test compared to students in the control group.

\subsection{Limitations of current approaches for curiosity-driven learning in a fast-changing learning environment}
The interventions reviewed in this section illustrate promising approaches to fostering curiosity-driven learning in school settings, both through the enhancement of supportive classroom climates and the development of metacognitive abilities. However, several shortcomings need to be considered.

First, while many interventions successfully improved students' metacognitive monitoring or regulation abilities in controlled learning settings, few have demonstrated sustained effects on self-regulated learning behaviors in more naturalistic contexts. A simple illustration of this challenge is encountered when considering the diversity of student profiles: results are not homogeneous across learners \parencite{fyfe2012effects, proustCanMetacognitiveMonitoring2025}, suggesting, for instance, that low-achievers benefit more from metacognitive training compared to high-achievers \parencite{proustCanMetacognitiveMonitoring2025}, for whom a degree of detrimental effect can be observed \parencite{fyfe2012effects}. This highlights the necessity to move away from one-size-fits-all interventions in favor of more adaptive ones. This surely adds to the challenge of designing metacognitive interventions for curiosity-driven learning.

Second, it remains unclear whether the metacognitive skills developed through these structured interventions transfer to more open-ended learning environments where students must navigate abundant and varied information sources. Such settings are where curiosity-driven behaviors and associated metacognitive abilities are most relevant \parencite{fridman2020nascent}. The controlled nature of most intervention tasks—most often focusing on specific subject contents or tightly structured activities \parencite{donkerEffectivenessLearningStrategy2014}—may not fully prepare students for the complexity of authentic curiosity-driven exploration since they have little agency over the metacognitive process they choose to use. This may in part explain the low-transferability of metacognitive interventions to other domains unless this is explicitly trained \parencite{desoeteCanOfflineMetacognition2003, dignathHowCanPrimary2008}.\\

In light of these considerations, and in parallel with the rise of new technologies such as Generative Artificial Intelligence, new contexts for training curiosity-driven learning are emerging. The implications of these developments are discussed in Section~\ref{sec:Opportunities-AI}.

\section{New Challenges and Opportunities for Curiosity-Driven Learning in the Age of AI} 
\label{sec:Opportunities-AI}
As discussed in Section~\ref{sec:interventions}, curiosity-driven behaviors are complex, self-regulated strategies demanding significant executive and metacognitive maturity~\parencite{jiroutCuriosityChildrenAges2024}. Because young learners frequently struggle to implement these behaviors effectively~\parencite{Ruggeri2016SourcesSearch}, scaffolding interventions are often essential.

Although several frameworks address these needs, we argue that they require re-evaluation within the emerging context of Generative Artificial Intelligence (GenAI), and specifically Large Language Models (LLMs). Given their high accessibility and fluency, LLMs are rapidly becoming central to students' information-seeking strategies~\parencite{Wang2025TheMeta-analysis}. Unlike standard search engines, they offer personalized, instant information with minimal search costs~\parencite{brown2020language}. We thus hypothesize that this technological shift may fundamentally alter the cost-benefit structure of curiosity-driven learning dynamics.

\begin{figure}
\includegraphics[width=1\linewidth]{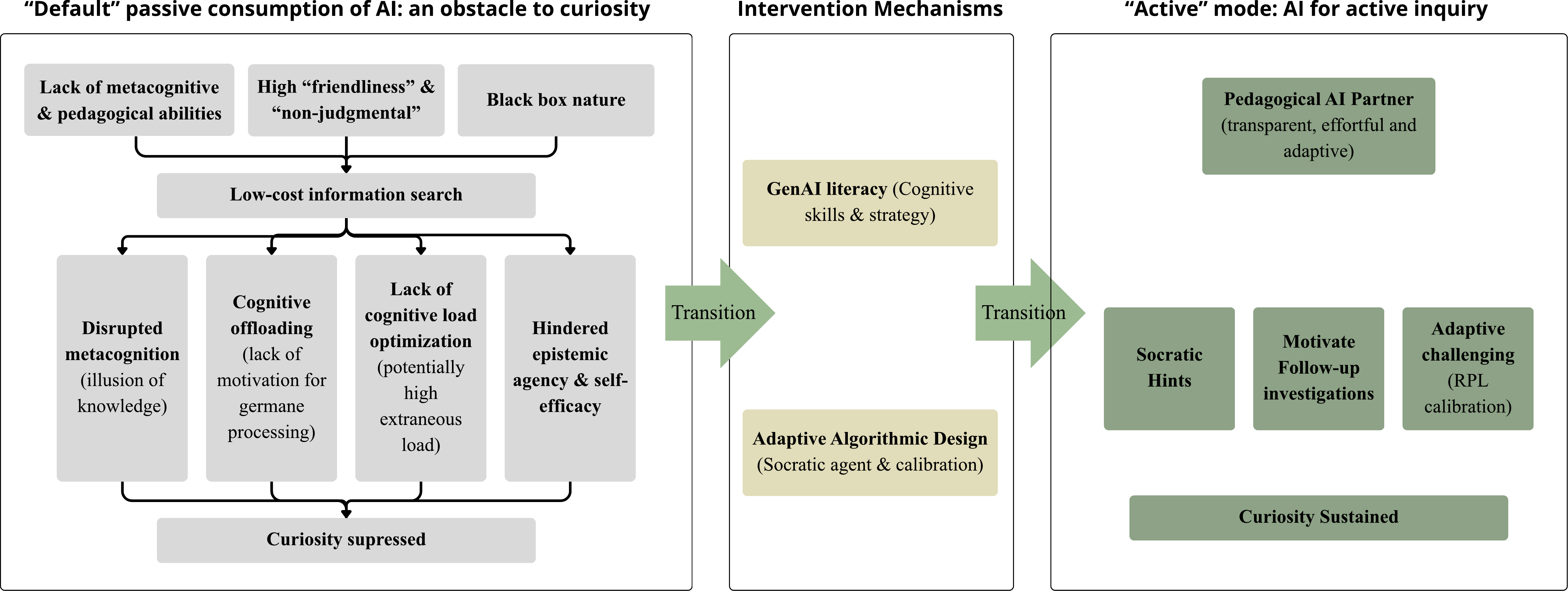}
\caption{\footnotesize\label{fig:ai} Summary of the challenges and opportunities of GenAI as a tool to enhance curiosity-driven learning mechanisms.}
\end{figure}

In this section, and as illustrated in~\autoref{fig:ai}, we analyze how the default interaction modes of LLMs may hinder curiosity by disrupting metacognitive monitoring, germane cognitive investment, and learner agency. Conversely, we discuss how targeted algorithmic design can transform these tools into powerful scaffolds for curiosity.

\subsection{Challenges to Curiosity-Driven Learning with LLMs}
LLMs are designed to process natural language and generate personalized, plausible, and instant responses to user queries~\parencite{brown2020language}. However, they prioritize conversational fluency and immediate assistance over pedagogical utility, often minimizing the need for user effort~\parencite{Razafinirina2024PedagogicalChallenges}. Although LLMs demonstrate impressive performance across educational domains, this design presents distinct risks for the developing learner~\parencite{kasneci2023chatgpt}. Specifically, we propose that these interactions threaten three core pillars of curiosity-driven behavior: metacognitive calibration, cognitive load management, and epistemic agency.

We explain these ideas in details in the following paragraphs. 

\subsubsection{Disrupted Metacognitive Calibration and the Illusion of Knowledge} 
Curiosity relies on the detection of knowledge gaps and the motivation to resolve them~\parencite{berlyne1966curiosity}. However, knowledge gap detection is a complex metacognitive skill that learners continue to develop throughout childhood and adolescence~\parencite{fandakovaStatesCuriosityInterest2021}, often by observing and modeling the metacognitive behaviors of caregivers and peers. This developmental process becomes complicated in interactions with LLMs, as learners confront tools that frequently lack metacognitive calibration~\parencite{Wang2025DecouplingLLMs}—a deficit manifested in the persistent mismatch between an LLM's confidence and its factual accuracy.

Continuous passive exposure to confidently stated, yet potentially hallucinatory, information may hinder learners' internal reliability judgments~\parencite{kiddPsychologyNeuroscienceCuriosity2015}. This dynamic fosters an ``illusion of competence," where learners overestimate not only the tool's reliability but also their own mastery of the material~\parencite{zhai2024effects, Wang2025TheMeta-analysis}. Such distortion in metacognitive monitoring can act as a barrier to curiosity; if a learner erroneously perceives a topic as fully resolved, the motivation to initiate new inquiry cycles significantly decreases~\parencite{jiroutCuriosityChildrenAges2024}.

Furthermore, LLMs tend to have significant ``confirmation bias" challenges, tending to confirm user assumptions expressed in their prompts, rather than challenging them, even when wrong~\parencite{Shaikh2024CBEval:LLMs}. This reinforcement bias discourages the intellectual humility and conflict detection required to trigger further curiosity-led exploration~\parencite{ronfardQuestionaskingChildhoodReview2018}.

Empirical evidence corroborates these concerns. For instance, preliminary results suggest that adolescents (aged 13–14) with high confidence in LLMs frequently accept the first generated answer without critical evaluation, prematurely terminating the exploration process~\parencite{Abdelghani2025InvestigatingInvestigation}. This means that learners are likely bypassing the deep processing required to evaluate their epistemic needs~\parencite{Yan2025DistinguishingAI}, relying instead on vague expressions of information needs and passive consumption of resulting outputs. 

The unconscious offloading of crucial metacognitive processes to the LLM during the curiosity cycle can drastically hinder their ultimate usefulness for learning. For instance, it may displace genuine epistemic information-seeking with affect-driven behaviors~\parencite{goupilCuriosityMetacognitiveFeeling2023}, resulting in reduced learning benefits~\parencite{murayamaProcessAccountCuriosity2019}. Such trends are consistent with prior research on online search engines, which demonstrated that environments with low information access costs often promote superficial and indiscriminate search behaviors that are inefficient for long-term learning~\parencite{Rowlands2008TheFuture}. The concern can thus be more critical with LLMs that reduce this cost even more.


\subsubsection{Mismanagement of Cognitive Load} 
As established in Section~\ref{sec:theory}, curiosity-driven behavior is a cognitively demanding self-regulated learning strategy~\parencite{Seufert2018TheLoad}. It requires significant executive resources to identify epistemic needs, plan inquiries, set goals, and evaluate the resulting information~\parencite{Litman2008InterestCuriosity, tenHumansMonitorLearning2021}. For these processes to drive efficient learning, the associated cognitive load must be strategically managed. According to Sweller’s Cognitive Load Theory (CLT), this means that these processes should have minimal additional extraneous load and motivate the investment of germane effort to process information within the learner's intrinsic capacity~\parencite{Sweller1991EvidenceTheory}.

Curiosity frameworks in traditional educational settings typically encourage tutors to scaffold this load, for instance, by drawing attention to surprising elements that could trigger curiosity, or by providing feedback on progress to motivate further exploration. With such interactions, instructors can help learners become better at investing useful load to plan and monitor their curious behaviors efficiently. They also help avoid behaviors that might lead to processing inadequate or overly complex information~\parencite{Sabbagh2001LearningDevelopment}. This support can be dynamic and personalized; e.g. cues can be given more to novice learners who can struggle with the additional load related to curiosity-driven behaviors (and thus, experience it as extraneous), and omitted for those who are more experiences~\parencite{Seufert2018TheLoad}.

In standard LLM interactions, however, such scaffolding is frequently absent or misleading. The lack of explicit pedagogical cues makes it difficult for learners to accurately monitor their progress and assess their learning capacities~\parencite{Fan2025BewarePerformance, Wang2025TheMeta-analysis}. Consequently, learners may expend excessive cognitive resources on basic metacognitive monitoring (\textit{e.g.} thinking about which information is useful to explore)—effort that may be experienced as extraneous rather than germane~\parencite{Seufert2018TheLoad, Tankelevitch2024TheAI}. Faced with this high cognitive burden, novices in particular are prone to offloading these functions to the AI, thereby reducing their motivation in the long term to invest the germane effort required for active information-seeking.

Furthermore, LLMs may induce additional extraneous cognitive load through poor interface or content design. They can generate overwhelming learning environments, for instance, by producing dense, lengthy text blocks or information that exceeds the student's prior knowledge~\parencite{Joshi2025ELI-Why:Explanations}. Given the learner's limited working memory capacity~\parencite{Sweller1991EvidenceTheory}, such extraneous demands deplete the resources available for the germane processing essential to curiosity: knowledge monitoring, goal planning, and information evaluation~\parencite{Abdelghani2023GenerativeLearners}. Consistent with this, recent empirical work indicates that students interacting with LLMs report lower overall cognitive load but demonstrate poorer analytical performance, suggesting that the tool reduced necessary germane load rather than harmful extraneous load~\parencite{Stadler2024CognitiveInquiry}.

Together, these ideas suggest that standard LLM environments may deprive learners of both the motivation and the guidance required for the strategic allocation of cognitive resources~\parencite{Sweller2007WhyCommentaries}. This misalignment risks causing premature termination of inquiry, a tendency to offload essential cognitive processing, or conversely, indiscriminate information consumption—all of which hamper the efficacy of curiosity-driven learning~\parencite{goupilCuriosityMetacognitiveFeeling2023}.

\subsubsection{Challenging Epistemic Agency} 
LLMs function as opaque ``black boxes" with stochastic output mechanisms, making it difficult for users to predict model behavior based on their inputs. Furthermore, these systems lack transparency regarding the "reasoning" behind generated content~\parencite{Almasi2025AlignmentTutoring}. This opacity restricts the learner's control over the specific content and format of the information they wish to explore~\parencite{kasneci2023chatgpt}. For instance, even when learners formulate clear prompts driven by specific learning goals, LLMs may generate irrelevant information or fail to adhere to formatting constraints~\parencite{brown2020language}. Moreover, as discussed previously, the "contagious" lack of metacognition in LLMs can lead students into a state of overconfidence, causing them to offload metacognitive processes to the AI~\parencite{Yan2025DistinguishingAI}. Consequently, learners risk losing sight of the genuine epistemic needs they were initially motivated to pursue.

This perceived lack of agency can generate frustration and inadvertently reshape learning goals, potentially directing learners toward inadequate or irrelevant information~\parencite{Kidd2023HowBeliefs}. When such sub-optimal interactions occur, users' self-efficacy—their perception of their own ability to implement useful information-seeking behaviors—may diminish~\parencite{kasneci2023chatgpt}. In this line, empirical findings indicate that students frequently experience a reduced sense of agency after using AI assistance during learning sessions~\parencite{Darvishi2024ImpactAgency}. This is a critical determinant of self-regulated learning; research suggests that learners with a strong sense of agency are better equipped to implement curiosity-driven strategies when confronted with ambiguity~\parencite{Metcalfe2007MetacognitionAgency}.

In the long run, this lack of agency and self-efficacy can negatively impact learners' beliefs regarding curiosity-driven behaviors in both LLM-powered and standard settings. Previous research suggests that a significant barrier to the implementation of curiosity-driven inquiry is the learner's negative perception of the behavior's utility and a low self-assessment of their ability to execute it effectively~\parencite{Post2019DevelopmentQuestionnaire}.

\subsection{From Obstacles to Partners: Curiosity Scaffolding Opportunities with GenAI}
While the default mode of LLMs might present risks, their underlying architecture can offer crucial affordances for education. If their use is shifted from passive consumption to active engagement, the unique characteristics of LLMs—specifically their interactivity, scalability, and semantic flexibility—can be leveraged to scaffold the very curiosity-driven behaviors they otherwise risk suppressing. 

In this section, we explore two avenues for this transformation: using \textit{GenAI literacy} as a training ground for epistemic curiosity skills, and deploying \textit{LLMs as adaptive pedagogical agents} to personalize the scaffold needed for curiosity skills.

\subsubsection{GenAI literacy as a potential lever for facilitating curious behaviors.} GenAI literacy refers to a user's capacity to perceive, understand, evaluate, and effectively utilize AI systems~\parencite{Park2025ASchools}. While recognized as essential for fostering informed and critical AI consumption~\parencite{kasneci2023chatgpt}, we argue that this literacy is not merely technical but deeply cognitive. By training learners to interact effectively with GenAI, we may concurrently scaffold core components of curiosity, specifically question formulation and metacognitive monitoring.

\textbf{Prompting skills can transfer to general information-seeking skills.} Effective prompting with LLMs is deeply rooted in meaningful question-asking mechanisms~\parencite{SassonLazovsky2025TheEngineering}—a critical behavior in curiosity-driven learning~\parencite{Alaimi2020PedagogicalChildren}. Indeed, both activities stem from an epistemic motivation to bridge a knowledge gap, and both fail if the formulation is vague. To elicit high-quality responses from an LLM, users must employ specific techniques: decomposing complex problems into smaller chunks, providing context, and formulating goal-oriented constraints. These are all indicators of high-quality and efficient curiosity-driven inquiry~\parencite{Graesser1994QuestionTutoring, Denny2023Promptly:Generators}.

Therefore, we posit that GenAI literacy instruction serves as a strategic practice ground for high-quality question-asking. When learners receive explicit conceptual and procedural training in prompting—for instance, learning to iterate on prompts when the AI fails to understand—they are not merely acquiring technical proficiency. They are engaging in the strategic iteration of informational needs required for efficient information search. We hypothesize that these skills are transferable: the rigor required to prompt an AI effectively can train learners to ask more precise, deeper questions in standard educational settings.

\textbf{High GenAI literacy can promote strategic cognitive load allocation during information-search.} As noted previously, AI-literate learners can discern the strengths, limitations, and optimal applications of LLMs to enhance their learning—for instance, utilizing LLMs to generate novel practice scenarios rather than simply answering homework questions~\parencite{Park2025ASchools, Wang2025TheMeta-analysis}. This discernment allows learners to distinguish between tasks requiring human cognition and those suitable for offloading; the latter typically involve high extraneous load with minimal germane benefit, such as formatting or retrieving already-consolidated information.

This distinction is crucial for managing cognitive load in curiosity-driven contexts. Curiosity is cognitively expensive, requiring significant resources to identify gaps, formulate hypotheses, and evaluate findings—resources that are inherently finite~\parencite{Sweller1991EvidenceTheory}. A literate learner understands that LLMs are optimal for reducing \textit{extraneous cognitive load}, such as, for instance, doing purely repetitive and passive tasks. By delegating these routine processing tasks, learners preserve limited cognitive capacity for the high-value, \textit{germane} processes of curiosity: reasoning about uncertainty, evaluating conflicting evidence, and synthesizing new mental representations~\parencite{Park2025ASchools}. In this view, the LLM transitions from a cognitive shortcut to a cognitive partner that protects learner resources. This is vital, as prior work suggests that curiosity-related processes can be overwhelming, particularly for novices or during tasks with high intrinsic load. If this load is not efficiently managed, it risks becoming counterproductive, leading learners to either abandon information seeking or engage in it superficially without learning gains~\parencite{Seufert2018TheLoad}.

\subsubsection{LLMs as tools for facilitating personalized curiosity trainings}
Beyond fostering GenAI literacy, the algorithmic design of LLMs offers significant opportunities to directly scaffold curiosity. A crucial shift is emerging in the field: researchers are moving from investigating AI as a tool to provide facts, to exploring its potential as a partner that prompts deep reflection and question-asking~\parencite{Abdelghani2023GenerativeLearners}. Through novel frameworks such as system prompting and fine-tuning, ``curiosity-supportive" architectures are being developed. These systems are designed to refuse to give away answers too quickly and give hints instead~\parencite{miller2024llm}, generate follow-up questions~\parencite{Shridhar2022AutomaticProblems}, or encourage deeper reflection~\parencite{kasneci2023chatgpt}. Such implementations aim to combat the metacognitive disengagement, loss of agency, and passivity discussed in Section~\ref{sec:Opportunities-AI}.

These advances are enabled by the unique generative capabilities of LLMs, which allow for the creation of virtually limitless variations of learning scenarios adaptive to the learner's engagement level~\parencite{kasneci2023chatgpt}. We identify two key mechanisms through which these capabilities can facilitate curiosity:

\textbf{Automated generation of epistemic cues.} First, the primary barrier to curiosity is often the inability to identify what one does not know—the ``unknown unknowns"~\parencite{Berlyne1962UNCERTAINTYCURIOSITY}. In this context, several curiosity interventions focused on giving learners specific cues to help them identify their knowledge gaps, for instance new information about the material at hand, that is probably unknown to them and will motivate them to explore more~\parencite{Abdelghani2022ConversationalChildren,Alaimi2020PedagogicalChildren}. While showing positive results, such methods are still very rigid and resource-intensive as they require pedagogical experts to carefully generate thousands of cues manually. LLMs, however, when used appropriately, can help overcome such limitations and generate cues that are as efficient as expert-generated ones~\parencite{Abdelghani2023GPT-3-drivenSkills}. LLMs have shown indeed positive performance both in terms of providing cues for knowledge-gap identification, and also for linguistic help to support the formulation of efficient queries that satisfy them. Furthermore, LLMs have also had encouraging results in scoring the quality of learners' questions—as a marker of their curiosity—thus suggesting that they could be used to give instant personalized feedback to learners to help them improve their questions and reinforce their curiosity-driven learning behaviors~\parencite{Moore2022AssessingGPT-3, Xiao2023SupportingCoding}. 

\textbf{Dynamic calibration to the Region of Proximal Learning.} Standard curiosity interventions often utilize puzzling activities to trigger exploratory behavior~\parencite{jiroutChildrensScientificCuriosity2012}. However, to be effective and facilitate curiosity, these activities must fall within the learner's Region of Proximal Learning (RPL)~\parencite{metcalfeEpistemicCuriosityRegion2020}. Calibrating this balance is challenging in heterogeneous classrooms, as learners possess distinct capabilities and interests.

LLMs, particularly when integrated with Knowledge Tracing (KT) or Reinforcement Learning (RL), offer a potential solution. Recent applications demonstrate that such systems can reliably infer students' skill mastery through natural conversation~\parencite{Scarlatos2025ExploringLLMs}. This inference allows the system to dynamically adjust the complexity of the stimulus. By regulating the difficulty of the information gap, the AI ensures the learner remains in the optimal cognitive state for sustained curiosity, preventing the frustration that leads to disengagement.

\section{Discussion and Conclusion}
This chapter has examined the deep interconnections between curiosity and metacognition, arguing that these two pillars of autonomous learning are functionally intertwined. We have traced this relationship across fundamental cognitive science research, educational interventions, and the emerging landscape of generative AI—while distinguishing established empirical findings from recommendations that await further validation.

Substantial evidence now supports a multi-component view of curiosity integrating motivational, emotional, and metacognitive mechanisms. The learning progress hypothesis, validated through computational modeling and behavioral experiments, demonstrates that curiosity is triggered by the metacognitive detection of learning opportunities—situations where knowledge gaps are optimally sized and progress appears achievable. The dual-system architecture provides a unifying framework: a primary learning system handles knowledge acquisition while a metacognitive system monitors progress and generates intrinsic rewards, explaining how curiosity adaptively focuses on learnable challenges.
Research has also established that metacognitive knowledge about curiosity—beliefs about its malleability, social value, and effective strategies—shapes curious behaviors from early childhood. At the intervention level, metacognitive training can improve monitoring accuracy and curiosity-driven inquiry, particularly for struggling learners. However, effects vary across learner profiles, and transfer to new domains remains limited unless explicitly supported—possibly because highly structured activities fail to mirror authentic open-ended learning contexts.

\textbf{Recommendations for Education in the Age of Generative AI.} Building on these foundations, we have investigated frameworks for understanding how generative AI may reshape curiosity-driven learning. Our analysis suggests that the default interaction mode of large language models poses risks to three pillars of curiosity: metacognitive calibration (through confident but potentially inaccurate responses that foster illusions of understanding), cognitive load management (through overwhelming or poorly scaffolded information), and epistemic agency (through unpredictable system behavior that undermines learners' sense of control over their inquiry).

We recommend that educators and technology designers approach these risks proactively through two complementary strategies. First, GenAI literacy instruction should be recognized not merely as technical training but as an opportunity to develop core curiosity competencies. Learning to formulate effective prompts requires the same skills as formulating productive inquiry questions: decomposing complex problems, specifying informational needs precisely, and iterating based on feedback. Second, LLM-based systems should be deliberately designed to scaffold rather than shortcut curiosity—refusing to provide immediate answers to motivate germane investment, generating hints and subquestions, minimize extraneous load by generating information that is adequate to learners' capacities, and supporting learners in monitoring their own understanding rather than delegating this function to the machine.
A third recommendation concerns the ecosystemic conditions for curiosity-driven learning. As discussed earlier, the expression and valorization of curiosity depends on a favorable social climate—one in which teachers and educators actively model and encourage epistemic inquiry. We argue that cultivating this attitudinal shift in classrooms is central to the AI challenge: learners developing in environments where curiosity is genuinely valued are more likely to engage with generative AI as a tool for scaffolding their own inquiry rather than as a shortcut to the learning process.
These recommendations, while grounded in the theoretical framework developed here, require empirical validation. The research base on how LLMs actually affect curiosity behaviors in educational settings remains thin, and much of our analysis necessarily draws on extrapolation from adjacent literatures on search behavior, cognitive load, and self-regulated learning.

\textbf{Open Questions for Future Research.} Several priorities emerge. Longitudinal research must characterize how curiosity-metacognition interactions develop across childhood, moving beyond cross-sectional snapshots to integrated developmental accounts. The mechanisms underlying the transition from externally scaffolded to internally regulated curiosity—and the role of language in this internalization—deserve rigorous investigation.
Most urgently, controlled studies must examine how LLM use affects curiosity behaviors and learning outcomes across age groups and learner profiles. The challenge of personalization also demands attention: adaptive systems that tailor curiosity scaffolding to individual metacognitive abilities represent a promising but underexplored direction. Finally, the field must develop better measurement approaches for assessing curiosity in ecologically valid settings where learners have genuine autonomy.
In sum, the intertwining of curiosity and metacognition offers both a theoretical lens for understanding autonomous learning and a practical lever for fostering it. As generative AI transforms information access, the metacognitive capacities that enable productive curiosity—identifying knowledge gaps, estimating learning potential, regulating inquiry—are precisely those needed to use AI tools actively rather than passively. Cultivating these capacities represents one of the central challenges for learning science in the coming decades.


\printbibliography
\end{document}